\documentclass[prb,twocolumn,amsmath,aps,pdffig,showpacs]{revtex4}
\usepackage{graphicx}
\usepackage{amsmath}
\usepackage{bm}
\usepackage{hyperref}
\usepackage{soul}
\usepackage{ulem}
\begin{document}
\title{Enhanced Andreev reflection in gapped graphene }

\author{Leyla Majidi and Malek Zareyan}

\affiliation{Institute for Advanced Studies in Basic Sciences
(IASBS), P. O. Box 45195-1159, Zanjan 45137-66731, Iran}

\begin{abstract}

We theoretically demonstrate unusual features of superconducting proximity effect in gapped graphene which presents a pseudospin symmetry-broken ferromagnet with a net pseudomagnetization. We find that the presence of a band gap makes the Andreev conductance of graphene superconductor/pseudoferromagnet (S/PF) junction to behave similar to that of a graphene ferromagnet-superconductor junction. The energy gap $\Delta_N$ can enhance the pseudospin inverted Andreev conductance of S/PF junction to reach a limiting maximum value for $\Delta_N\gg \mu$, which depending on the bias voltage can be larger than the value for the corresponding junction with no energy gap. We further demonstrate a damped-oscillatory behavior for the local density of states of the PF region of S/PF junction and a long-range crossed Andreev reflection process in PF/S/PF structure with antiparallel alignment of pseudomagnetizations of PFs, which confirm that, in this respect, the gapped normal graphene behaves like a ferromagnetic graphene.
\end{abstract}


\maketitle

\section{\label{sec:intro}Introduction}
Transmission of low energy electrons through a normal-metallic-superconducting (N/S) junction is realized via Andreev reflection (AR)\cite{Andreev64} through which an electron with an energy $\varepsilon$ (relative to the chemical potential $\mu$) and spin polarization $\sigma$, upon hitting the N/S interface, is retro reflected as a hole with the same energy but opposite spin direction $-\sigma$.  This peculiar scattering process results in a finite conductance of a N/S junction at the bias voltages below the superconducting energy gap $eV<\Delta_S$\cite{Blonder82}. When a ferromagnetic (F) metal is interfaced to a superconductor, the exchange splitting of up and down spin subbands leads to a suppression of AR at F/S interface. As the result, the subgap Andreev conductance decreases with increasing the exchange energy $h$ from its value for N/S junction and vanishes for a half metallic ferromagnet with $h=\mu$, where all carriers have the same spin\cite{de Jong95}. The exchange field causes a momentum shift of order $2h/v_F$  between Andreev correlated electron-hole in F metal, that is responsible for the spatial damped oscillations of the induced superconducting correlations in F metals\cite{Kontos01,Zareyan01,Zareyan02,Buzdin05}.
\par
Novel interesting phenomena arise when N/S and F/S proximity structures are realized in graphene, the recently discovered two dimensional solid of carbon atoms with honeycomb lattice structure\cite{Novoselov04,Novoselov05,Zhang05}. Graphene has a zero-gap semiconducting band structure in which the charge carriers behave like 2D massless Dirac fermions with a pseudo-relativistic chiral property. The carrier type, [electron-like ($n$) or hole-like ($p$)] and its density can be tuned by means of electrical gate or doping of underlying substrate. Recent experimental progresses in proximity-inducing superconductivity in graphene by fabrication of transparent contacts between a graphene monolayer and a superconductor [see for instance, Refs. \onlinecite{Heersche07,Du08,Jeong11}], has provided a unique possibility to study relativistic-like superconductivity and proximity effect. Peculiarity of AR in graphene N/S junctions has been explained by Beenakker, who predicted the possibility for a specular AR in undoped normal graphene, and its associated anomaly in Andreev current-voltage characteristics of a graphene N/S contact\cite{beenakker06,beenakker08}. In the case of a graphene F/S junction, the situation is dramatically different from common F/S junctions. It has been shown that for the exchange energies higher than the chemical potential $h>\mu$, a peculiar spin-resolved Andreev-Klein process at graphene F/S interface can result in an enhancement of the subgap Andreev conductance by $h$, up to the point at which the conductance at low voltages $eV\ll\Delta_S$ is larger than its value for the corresponding N/S structure\cite{Zareyan08,Asano08,Zhang08}. Also, the corresponding Andreev-Klein bound states in graphene S/F/S structure are responsible for the long-range Josephson coupling of F graphene\cite{Moghaddam08,Linder08}. Moreover, specific non-local proximity effect takes place in graphene-based superconducting heterostructures mediating purely by a non-local process known as crossed Andreev reflection (CAR) which creates a spatially entangled electron-hole pair. While in ordinary non-relativistic systems the small value of CAR conductance is canceled by the conductance of elastic electron co-tunneling (CT) process, it can be enhanced in ballistic graphene N/S/N and F/S/F structures\cite{Cayssol08,linder09}.
\par
The above explained unusual properties of AR in graphene arise from its electronic structure which is fundamentally different from that of a metal or semiconductor. Further peculiarity of graphene comes from the fact that its electrons, in addition to the regular spin, appear to come endowed with the two quantum degrees of freedom, the so called pseudospin and valley. The pseudospin represents the sublattice degree of freedom of the graphene's honeycomb structure, and the valley defines the corresponding degree of freedom in the reciprocal lattice\cite{Novoselov05,Wallace47,Slonczewski58,Haldane88,Castro09}. In this paper, we demonstrate still another peculiarity of graphene-based superconducting hybrid structures which is resulted from the pseudospin degree of freedom of electrons in graphene with a (non-superconducting) gap in its electronic spectrum. The effect of the pseudospin and the valley  degrees of freedom has already been proven to be drastically important in several quantum transport phenomena in graphene\cite{Novoselov05,Zhang05,Gusynin05,Du09,Katsnelson06,Young09,Stander09,Peres06,Tworzydlo06,DiCarlo08}, such that these additional quantum numbers have been proposed to be used for the controlling electronic devices in  the same way as the electron spin which is used in spintronics and quantum computing\cite{Rycerz07,Akhmerov08,Wu11}. Control of the pseudospin of the electron has been explained to be more feasible in both monolayer graphene and bilayer graphene\cite{Min08,Jose09,Trushin11}. In particular, it was shown that the monolayer graphene with a gap in its electronic spectrum and an appropriate doping presents a pseudospin symmetry-broken ferromagnet, with a finite pseudospin magnetization oriented vertically to the graphene plane\cite{majidi11}. Here, we find that the pseudospin degree of freedom has a determining effect on AR and the associated proximity effect in hybrid structures of gapped normal graphene regions as pseudoferromagnets (PFs) and a superconductor in S/PF and PF/S/PF geometries.
\par
 In graphene PF, due to the possibility for a small chemical potential $\mu$, an electron from the conduction-band can be reflected as a hole in the valance-band, depending on the electron energy $\varepsilon$, $\mu$ and the energy gap $\Delta_N$. In terms of the carriers pseudospin vector, this corresponds to an inversion of the $z$ component of the pseudospin vector. Thus, such a peculiar AR is associated with a transition from the $n$-type carriers to the $p$-type carriers, which is called Klein tunneling\cite{Katsnelson06,Young09,Stander09} in analogous to the corresponding effect in relativistic quantum mechanics\cite{Klein29}. We obtain that such a Andreev-Klein process at graphene S/PF junction can enhance the amplitude of AR and the resulting Andreev conductance by $\Delta_N$. In particular, we show that depending on the bias voltage the Andreev conductance of weekly doped PF ($\mu\ll\Delta_N$) can be larger than its value for the corresponding graphene S/N junction. We also demonstrate that depending on the energy $\varepsilon$ of the incident electron, $\mu$ and $\Delta_N$, AR can be of retro or specular types, respectively, without or with the inversion of the $z$ component of the pseudospin vector. We further demonstrate that the local density of states (DOS) inside PF has a damped-oscillatory behavior which approves that the energy gap $\Delta_N$ in the band structure of normal graphene produces an effect similar to the exchange field in F graphene. For PF/S/PF junction system, we find that the transport is mediated purely by CT process in parallel alignment of pseudomagnetizations (PMs) and CAR process in antiparallel configuration, that is accompanied by pseudospin switching effect and confirms the crucial rule of the pseudospin in the gapped normal graphene and its similarity to the rule of spin in a F graphene.
\par
This paper is organized as follows. In Sec.\ \ref{sec:level1}, we establish the theoretical framework which will be used to investigate AR in graphene S/PF junction. We present our main findings for the Andreev conductance and the proximity DOS of the S/PF junction, respectively, in Secs. \ \ref{sec:level2} and \ref{sec:level3}. Section \ \ref{sec:level4} is devoted to the investigation of CAR in PF/S/PF junction. Finally, we present the conclusion in Sec. \ \ref{sec:level5}.

\section{\label{sec:level1}Model and Basic Equations}
\begin{figure}
\begin{center}
\includegraphics[width=3.4in]{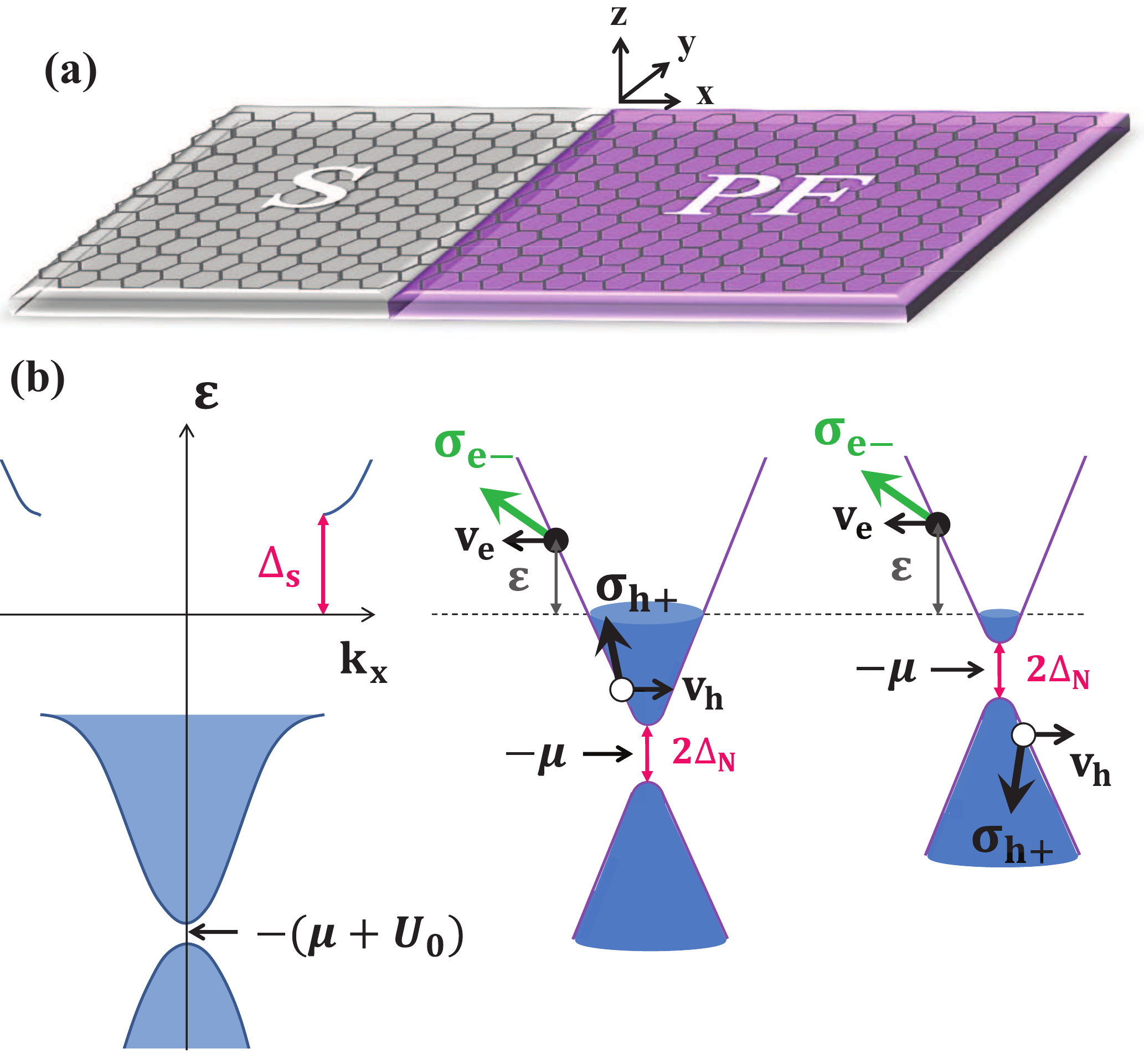}
\end{center}
\caption{\label{Fig:1}(Color online)(a) Schematic illustration of the graphene S/PF junction. (b) The energy band diagram of the S and PF regions to explain the two cases of Andreev reflection at S/PF interface. Left panel: the energy-momentum relation of the highly doped S graphene. Middle and right panels: the band structure of the n-doped PF region for two values of the chemical potential. Middle (Right) panel shows that an incident electron from the conduction-band of PF region with a subgap energy $0\leq\varepsilon\leq \mu-\Delta_N$ ($\varepsilon\geq \mu+\Delta_N$) is reflected as a hole in the conduction- (valance-)band without (with) the inversion of the $z$ component of the pseudospin at S/PF interface. $\bm{\sigma}_{e-}$ and $\bm{\sigma}_{h+}$ denote the pseudospin vectors of incident electron and reflected hole. $v_{e}$ and $v_{h}$ denote the velocity vectors of the electron and the hole, moving in different directions.}
\end{figure}
We consider a wide graphene S/PF junction normal to $x$-axis with highly doped superconducting (S) region for $x<0$ and n-doped pseudoferromagnetic (PF) region  for $x>0$ [see Fig. \ref{Fig:1}(a)]. Pseudoferromagnetic region is a monolayer graphene with an energy gap $\Delta_N$ in its electronic band structure, which behaves as a pseudospin symmetry-broken ferromagnet with a perpendicular to the plane of graphene pseudomagnetization (PM) whose direction is switched by altering the type of doping between $n$ and $p$. The magnitude of PM depends on the ratio of the chemical potential to the energy gap $\mu/\Delta_N$, such that the vertical pseudomagnetization per electron $PM_z/N$ takes its maximum value $PM_z/N = 1$ for $\mu\simeq\Delta_N$, decreases by increasing $\mu/\Delta_N$ and goes to zero for highly doped gapped graphene ($\mu\gg\Delta_N$)\cite{majidi11}. The most common practical approach to induce a band gap is to realize graphene on top of the appropriate substrate like SiC which breaks the sublattice symmetry\cite{Zhou07,Varchon07}.
The S part can be produced by depositing S electrode on top of the graphene sheet\cite{Heersche07,Du08,Jeong11}.
In this region $\Delta_N=0$ and the superconducting correlations are characterized by the superconducting pair potential (order parameter) $\Delta_S$ which is taken to be real and constant. To study AR at S/PF interface within the scattering formalism, we first construct the quasiparticle wave functions that participate in the scattering processes. In order to describe the superconducting correlations between relativistic electrons and holes of different valleys, we adopt Dirac-Bogoliubov-de Gennes (DBdG) equation\cite{beenakker06} which has the form
\begin{equation}
\label{DBdG}
\hspace{-0.5cm}\left(
\begin{array}{cc}
H-\mu & \Delta_S \\
\Delta_{S}^{\ast}& \mu-H
\\
\end{array}
\right)
\left(
\begin{array}{c}
u\\
v
\end{array}
\right)
=\varepsilon\left(
\begin{array}{c}
u\\
v
\end{array}
\right),
\end{equation}
where
\begin{equation}
\label{H}
H=v_{F}(\bm{\sigma}.\bm{p})+\Delta_N \sigma_{z}-U(\bm{r}),
\end{equation}
is the two-dimensional Dirac Hamiltonian with an energy gap, $\varepsilon$ is the excitation energy and $U(\bm{r})$ the electrostatic potential is taken to be $U_0\gg \mu$ in S region and $U=0$ in PF region. The electron and the hole wave functions, $u$ and $v$, are two-component spinors of the form $(\psi_A,\psi_B)$ and $\bm{\sigma}=(\sigma_x,\sigma_y,\sigma_z)$
is the vector of the Pauli matrices operating in the space of the two sublattices of the honeycomb lattice structure.
\par
 An incident electron of the conduction-band from right to S/PF interface with a subgap energy $\varepsilon\leq\Delta_S$ can be either normally reflected as an electron in the conduction-band or Andreev reflected as a hole in the conduction- or the valance-band. As illustrated in Fig. \ref{Fig:1}(b), as long as $0\leq\varepsilon\leq \mu-\Delta_N$ the reflected hole is an empty state in the conduction-band and AR is retro (middle panel), while for $\varepsilon\geq \mu+\Delta_N$ it is an empty state in the valance-band and AR is specular, if $\Delta_N<\Delta_S$ (right panel). The left panel of Fig. \ref{Fig:1}(b) shows the energy-momentum relation of the highly doped S graphene \cite{beenakker06}. The importance of AR near the Fermi level imposes the condition of $\Delta_N<\Delta_S$ on size of the energy gap $\Delta_N$. The retro reflection dominates if $\mu\gg\Delta_S+\Delta_N$, while the specular reflection dominates if $\mu\ll\Delta_S-\Delta_N$. Using the same method as in Ref. \onlinecite{majidi11}, the pseudospin of the incident electron and the reflected hole of the conduction- (valance-)band are obtained as,
\begin{eqnarray}
\label{pseudospin_e}
&&\hspace{-5mm}\langle\bm{\sigma}(\bm{k})\rangle_{c}^{e-}=\sqrt{1-(\frac{\Delta_N}{\mu+\varepsilon})^2}\ (-\cos{\alpha_e}\ \hat{x}+\sin{\alpha_e}\ \hat{y})\nonumber\\
&&\hspace{1.4cm}+\frac{\Delta_N}{\mu+\varepsilon}\ \hat{z},\\
\label{pseudospin_h}
&&\hspace{-5mm}\langle\bm{\sigma}(\bm{k})\rangle_{c(v)}^{h+}=\sqrt{1-(\frac{\Delta_N}{\mu-\varepsilon})^2}\ (-\cos{\alpha_h}\ \hat{x}\pm\sin{\alpha_h}\ \hat{y})\nonumber\\
&&\hspace{1.5cm}\pm\frac{\Delta_N}{|\mu-\varepsilon|}\ \hat{z},
\end{eqnarray}
where
\begin{equation}
\alpha_{e(h)}=\arcsin{(\frac{\hbar vq}{\sqrt{(\mu\pm\varepsilon)^2-{\Delta_N}^2}})}
\end{equation}
indicates the angle of propagation of the electron (hole), at a transverse momentum $q$ with energy-momentum relation
\begin{equation}
\varepsilon_{c}^e=-\mu+\sqrt{{\Delta_N}^2+(\hbar v|{\bm{k}}_{e}|)^2}
\end{equation}
for the conduction-band electron and
\begin{equation}
\varepsilon_{c(v)}^h=\mu\mp\sqrt{{\Delta_N}^2+(\hbar v|{\bm{k}}_{h}|)^2}
\end{equation}
 for the hole from the conduction- (valance-)band. Here, the superscripts $\pm$ denote the two directions of propagation along the $x$-axis.
 As can be seen from the above equations when an electron from the conduction-band is reflected as a hole in the valance-band, the sign of the gap-induced $z$ component of the pseudospin vector $\langle{\sigma}_z\rangle$ is changed, while in the case of the conduction-band hole, it retains its sign.
 This is shown schematically in the middle (right) panel of Fig. \ref{Fig:1}(b) for the case of Andreev reflected hole from the conduction- (valance-)band without (with) the inversion of $\langle{\sigma}_z\rangle$ upon AR at S/PF interface.
 Thus, for the incident electron and the reflected hole being from different types of bands, we have an inversion of the $z$ component of the pseudospin vector upon AR at S/PF interface. In the following we will show how the pseudospin $\langle{\sigma}_z\rangle$ inversion by AR leads to peculiar properties of S/PF and PF/S/PF systems.
\par
Denoting the amplitudes of normal and Andreev reflections, respectively, by $r_{c(v)}$ and $r_{A,c(v)}$, the wave functions inside PF and S regions are written as
\begin{eqnarray}
\label{pf wave function}
&&\psi_{PF}^{c(v)}=\psi_{c}^{e-}+r_{c(v)}\ \psi_c^{e+}+r_{A,c(v)}\ \psi_{c(v)}^{h+},\\
\label{s wave function}
&&\psi_{S}=a\ \psi^{S+}+b\ \psi^{S-},
\end{eqnarray}
where $\psi_{c(v)}^{e(h)\pm}$ and $\psi^{S\pm}$ are the solutions of DBdG equation for the quasiparticles inside PF and S regions, respectively,
and the two cases of the Andreev reflected holes from the conduction- (valance-)band without (with) the inversion of $\langle{\sigma}_z\rangle$ are denoted by $c(v)$ in $\psi_{PF}^{c(v)}$.
Matching the wave functions of PF and S regions at the interface $x=0$, we obtain the normal and Andreev reflections amplitudes as
\begin{eqnarray}
\label{r}
&&\hspace{-7mm}r_{c(v)}={\frac{a_{c(v)}-e^{i\alpha_e+\phi_e}+e^{i\alpha_h+\phi_h}[1-a_{c(v)}e^{i\alpha_e+\phi_e}]}
{a_{c(v)}e^{i\alpha_e}+e^{\phi_e}+e^{i\alpha_h+\phi_h}[a_{c(v)}e^{\phi_e}+e^{i\alpha_e}]}},\\
\label{rA}
&&\hspace{-7mm}r_{A,c(v)}=\frac{b_{c(v)}}{\sqrt{\cos{\alpha_e}\cos{\alpha_h}}}\ e^{(\phi_e+\phi_h)/2}\nonumber\\
&&\hspace{5mm}\frac{e^{-i(\alpha_e-\alpha_h)/2}\ (1+e^{2i\alpha_h})}{a_{c(v)}e^{i\alpha_e}+e^{\phi_e}+e^{i\alpha_h+\phi_h}[a_{c(v)} e^{\phi_e}+e^{i\alpha_e}]},
\end{eqnarray}
where
\begin{eqnarray}
\phi_{e(h)}&=& \operatorname{arcsinh}{(\frac{\Delta_N}{\sqrt{(\mu\pm\varepsilon)^2-{\Delta_N}^2}})},\\
\hspace{4mm}&&\beta=\arccos{(\varepsilon/\Delta_S)},
\end{eqnarray}
$a_{c(v)}={(i\tan{\beta})}^{\pm1}$ and $b_{c(v)}=\sec{\beta}\ (-i\csc{\beta})$.
Using the above-found reflection amplitudes and wave functions, we investigate the Andreev conductance of S/PF interface
and the proximity DOS inside the PF region, respectively, in the next two sections.
\section{\label{sec:level2}Andreev conductance}
In this section, we evaluate the Andreev conductance of S/PF junction by using Blonder-Tinkham-Klapwijk (BTK) formula\cite{Blonder82},
\begin{eqnarray}
\label{G}
&&\hspace{-10mm}G_{c(v)}=G_0\int_{0}^{\alpha_{c}}(1-|r_{c(v)}|^2+|r_{A,c(v)}|^2)\cos\alpha_e\ d\alpha_e,\\
&&\hspace{-10mm}G_0=\frac{4e^2}{h}\tilde{N}(eV),\ \tilde{N}(\varepsilon)=\frac{W(\mu+\varepsilon)^2}{\pi\hbar v_F\sqrt{(\mu+\varepsilon)^2-{\Delta_N}^2}},
\end{eqnarray}
where we put $\varepsilon=eV$ at zero temperature. The quantity $G_0$ is the ballistic conductance of $\tilde{N}$ transverse modes in a sheet of gapped graphene of width W and
\begin{equation}
\alpha_{c}=\arcsin{(\sqrt{\frac{{(\mu-\varepsilon)^2-{\Delta_N}^2}}{{(\mu+\varepsilon)^2-{\Delta_N}^2}}})}
\end{equation}
is the critical angle of incidence above which the Andreev reflected waves become evanescent and do not contribute to any transport of charge.
\begin{figure}
\begin{center}
\includegraphics[width=3.3in]{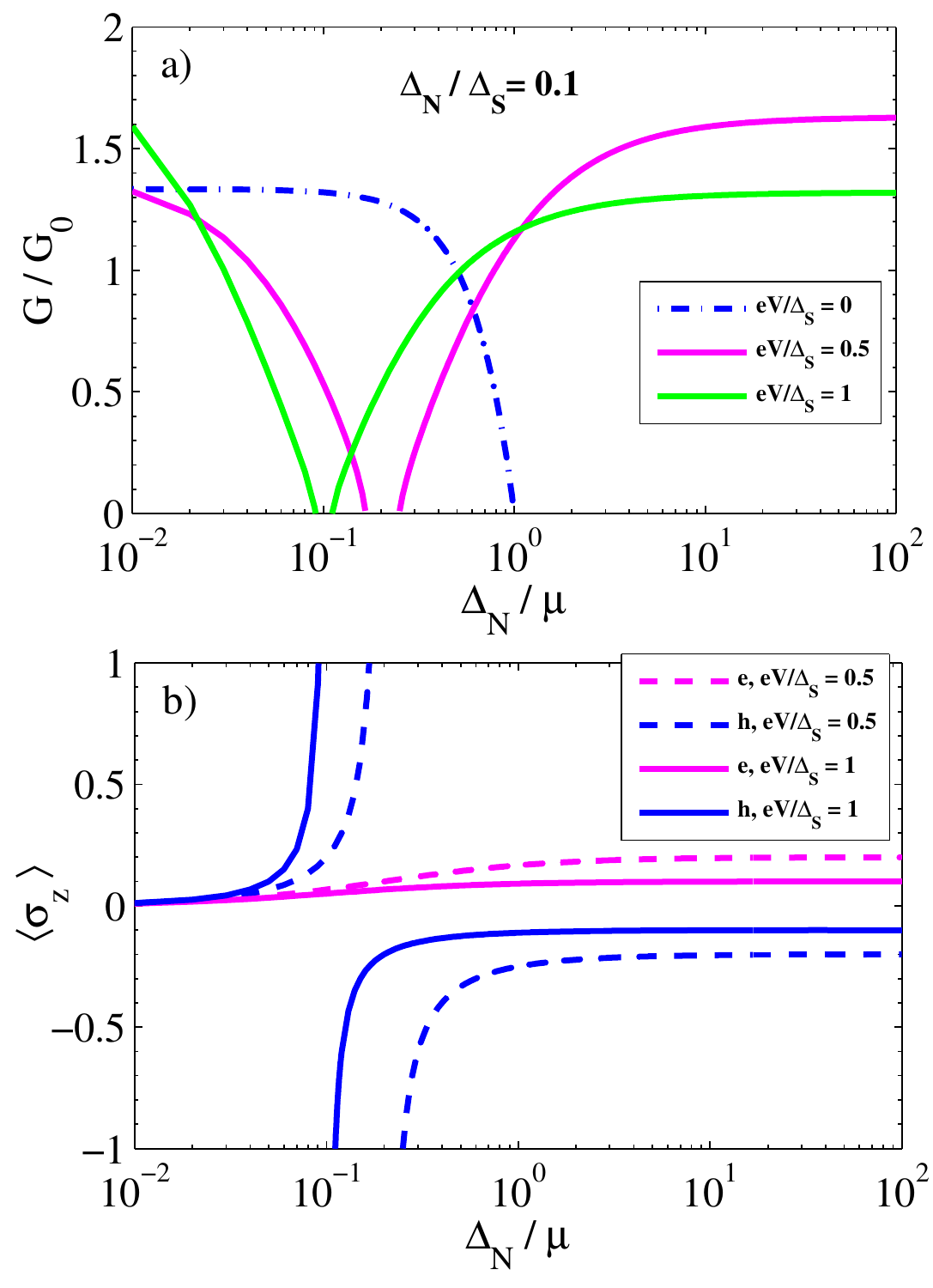}
\end{center}
\caption{\label{Fig:2}(Color online) (a) Dependence of the Andreev conductance of graphene S/PF contact on the gap $\Delta_N/\mu$ (in units of the chemical potential) at three bias voltages $eV/\Delta_S = 0,0.5,1$. (b) The behavior of the $z$ component of the pseudospin vector $\langle{\sigma}_z\rangle$ versus $\Delta_N/\mu$ for the incident electron from the conduction-band and the reflected hole from the conduction (before the gap) or valance-band (after the gap) at two values of the bias voltage $eV/\Delta_S = 0.5,1$, when $\Delta_N / \Delta_S=0.1$.}
\end{figure}
\par
Dependence of the resulting Andreev conductance $G/G_0$ on the ratio $\Delta_N/\mu$ is presented in Fig. \ref{Fig:2}(a) for $\Delta_N/\Delta_S=0.1$ and three different bias voltages $eV/\Delta_S = 0,0.5,1$. For $\Delta_N<{\mu/(1+eV/\Delta_N)}$, the conductance decreases monotonically with $\Delta_N/\mu$. In this interval, the incident electron and the reflected hole are from the conduction-band and therefore AR is without the inversion of $\langle{\sigma}_z\rangle$ [see Fig. \ref{Fig:2}(b)]. The density of states of the conduction-band hole decreases by increasing $\Delta_N/\mu$. Thus, the amplitude of AR and hence the Andreev conductance decreases with $\Delta_N/\mu$ and goes to zero at $\Delta_N={\mu/(1+eV/\Delta_N)}$, where the density of states of the conduction-band hole vanishes. There is a gap in conductance for ${\mu/(1+eV/\Delta_N)}<\Delta_N<{\mu/(eV/\Delta_N-1)}$, which decreases with $eV/\Delta_S$ and goes towards smaller $\Delta_N/\mu$. For $\Delta_N\geq{\mu/(eV/\Delta_N-1)}$, the pseudospin $\langle{\sigma}_z\rangle$ inverted Andreev conductance increases monotonically with $\Delta_N/\mu$. In this regime, the transport is between the conduction and the valance-band and the incident electron of the conduction-band is reflected as a hole in the valance band. So the pseudospin $\langle{\sigma}_z\rangle$ of the reflected hole changes sign [see Fig. \ref{Fig:2}(b)] and the density of states of the hole increases with $\Delta_N/\mu$, resulting in an enhancing Andreev conductance. In this case, the pseudospin $\langle{\sigma}_z\rangle$ inverted AR is associated with a Klein tunneling of the $n$-type carriers to the $p$-type carriers. The enhancing conductance reaches a limiting maximum value which depends on the bias voltage. Importantly we see that depending on the value of the bias voltage, the limiting value of $G/G_0$ for $\Delta_N\gg \mu$ can be larger than its value for the corresponding S/N structure ($\Delta_N\ll \mu$).
\par
\begin{figure}[htb]
\begin{center}
\includegraphics[width=3.2in]{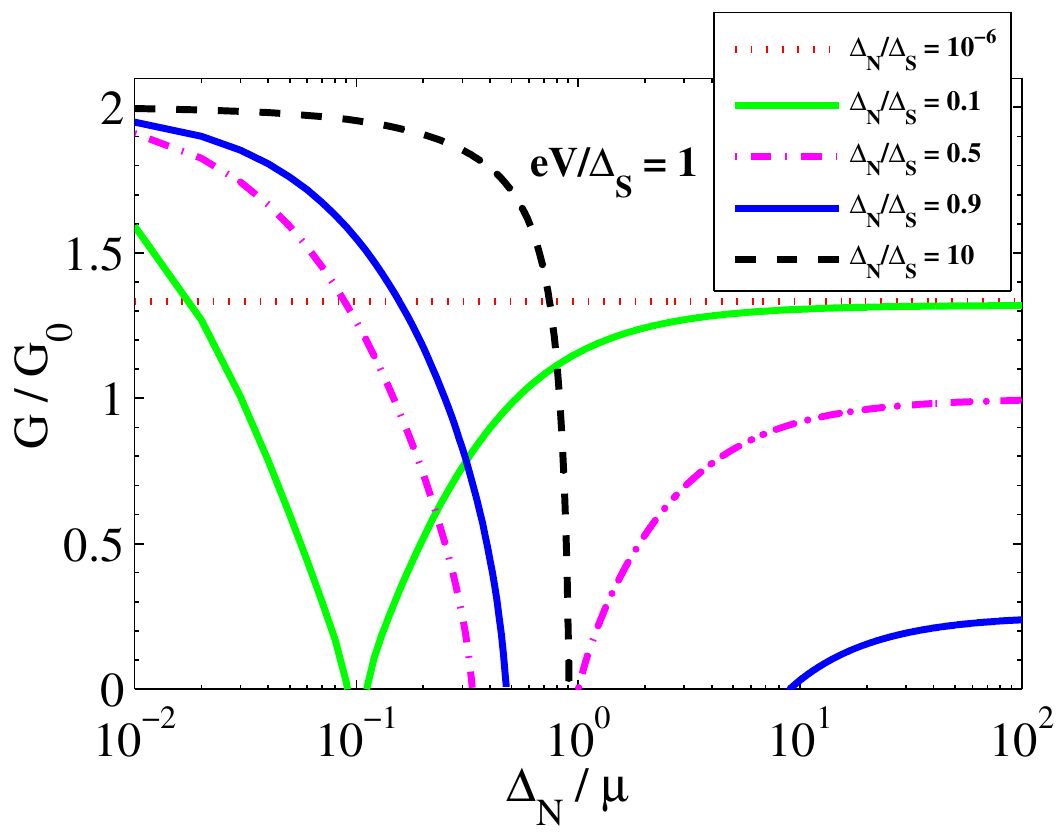}
\end{center}
\caption{\label{Fig:3}(Color online) Plot of the Andreev conductance versus $\Delta_N/\mu$ for different values of $\Delta_N/\Delta_S$, when $eV/\Delta_S = 1$.}
\end{figure}
For $eV/\Delta_S = 0$, we only have AR without $\langle{\sigma}_z\rangle$ inversion, while for finite bias voltages AR without the inversion of $\langle{\sigma}_z\rangle$ changes to the inverting type when $\Delta_N/\mu$ goes across the conductance gap.  In order to explain the behavior of the carriers pseudospin vector in AR process, we have plotted the $\Delta_N/\mu$ dependence of the $z$ component of the pseudospin vector $\langle{\sigma}_z\rangle$ for the incident electron $e$ and the reflected hole $h$ in Fig. \ref{Fig:2}(b) when $eV/\Delta_S = 0.5,1$. In the limit of $\Delta_N\ll \mu$, $\langle{\sigma}_z\rangle$ of the electron and the hole are zero and the system behaves like a graphene S/N structure. Increasing $\Delta_N/\mu$ leads to the out of plane component for the pseudospin vector of the electron and the hole such that $\langle{\sigma}_z\rangle$ of the hole increases more than that of the electron. In this regime, the conductance decreases monotonically with $\Delta_N/\mu$ and goes to zero at $\Delta_N={\mu/(1+eV/\Delta_N)}$. As is shown in Fig. \ref{Fig:2}(b), the absence of hole states inside the PF gap causes a gap in conductance. For $\Delta_N\geq {\mu/(eV/\Delta_N-1)}$, $\langle{\sigma}_z\rangle$ of the hole changes sign and decreases with $\Delta_N/\mu$. In this case, the Andreev conductance increases monotonically from zero and reaches a limiting maximum value for $\Delta_N\gg \mu$ (the limit of specular AR), where $\langle{\sigma}_z\rangle$ of the electron and the hole have equal magnitudes and different signs. According to Eq. (\ref{pseudospin_e}), the magnitude of $\langle{\sigma}_z\rangle$ for the electron decreases with $eV/\Delta_S$ such that in the limit of $\Delta_N\gg \mu$, it reaches $0.2$ and $0.1$ for $eV/\Delta_S = 0.5$ and $1$, respectively.
\par
Also the behavior of the Andreev conductance versus $\Delta_N/\mu$ is shown in Fig. \ref{Fig:3} for different values of $\Delta_N/\Delta_S$, when $eV/\Delta_S = 1$. It is seen that the limiting value of $G/G_0$ increases by increasing $\Delta_N/\Delta_S$ for $\Delta_N\ll \mu$ and tends to the corresponding value of a retro type AR $(G/G_0 = 2)$ as $\Delta_N\rightarrow\Delta_S$, while for $\Delta_N\gg \mu$ it decreases from its value for a specular AR in corresponding S/N structure $\Delta_N\ll\Delta_S$ $(G/G_0 = 4/3)$ and vanishes for $\Delta_N>\Delta_S$. Also the Andreev conductance gap is getting broadened by increasing $\Delta_N/\Delta_S$.
\par
So the behavior of the Andreev conductance with $\Delta_N/\mu$ is similar to that of a graphene F/S junction with $h/\mu$, where AR of n-n type carriers for $h<\mu$ changes to the Andreev-Klein reflection of the n-p type carriers for $h>\mu$\cite{Zareyan08}. This tells us that the energy gap $\Delta_N$ in the band structure of normal graphene behaves like an exchange energy in F graphene and enhances
 the subgap Andreev conductance of S/PF junction, which is accompanied by the inversion of the $z$ component of the pseudospin vector for the reflected hole relative to the incident electron.
\section{\label{sec:level3}Local Density of States}
Let us now study the proximity effect in S/PF junction by focusing on an experimentally accessible quantity, the local density of states (DOS)
inside the PF region. To find the proximity DOS as a function of energy and position, we use the formula \cite{Gennes89}
\begin{equation}
\label{N}
N(\varepsilon,r)=\sum_{\bm{k}}{|\psi_{\bm{k}}(r)|^2\ \delta(\varepsilon(\bm{k})-\varepsilon)},\\
\end{equation}
where $\psi_{\bm{k}}(r)$ corresponds to the eigenfunction of energy $\varepsilon(\bm{k})$ and the sum is over all states with the wave vectors  $\bm{k}$. Replacing the wave function of Eq. (\ref{pf wave function}) in the above equation, we find the total subgap DOS inside the PF region for two cases of AR without (with)
$\langle{\sigma}_z\rangle$ inversion as,
\begin{eqnarray}
&&\hspace{-20mm}\frac{N^{c(v)}(\varepsilon, x)}{N_{0}(\varepsilon)} =\frac{1}{4}\sqrt{1- (\frac{\Delta_N}{\mu+\varepsilon})^2}\nonumber\\
&&\int_{-\pi/2}^{\pi/2} |{\psi(r)}_{PF}^{c(v)}|^2\ {\cos}^2\alpha_e\ d\alpha_e,\\
N_0(\varepsilon)&=& \frac{(\mu+\varepsilon)^2}{(\pi\hbar v_F)^2 \sqrt{(\mu+\varepsilon)^2-{\Delta_N}^2}}.
\end{eqnarray}
Here, $N_0(\varepsilon)$ is the DOS of a PF layer.
\begin{figure}
\begin{center}
\includegraphics[width=3.3in]{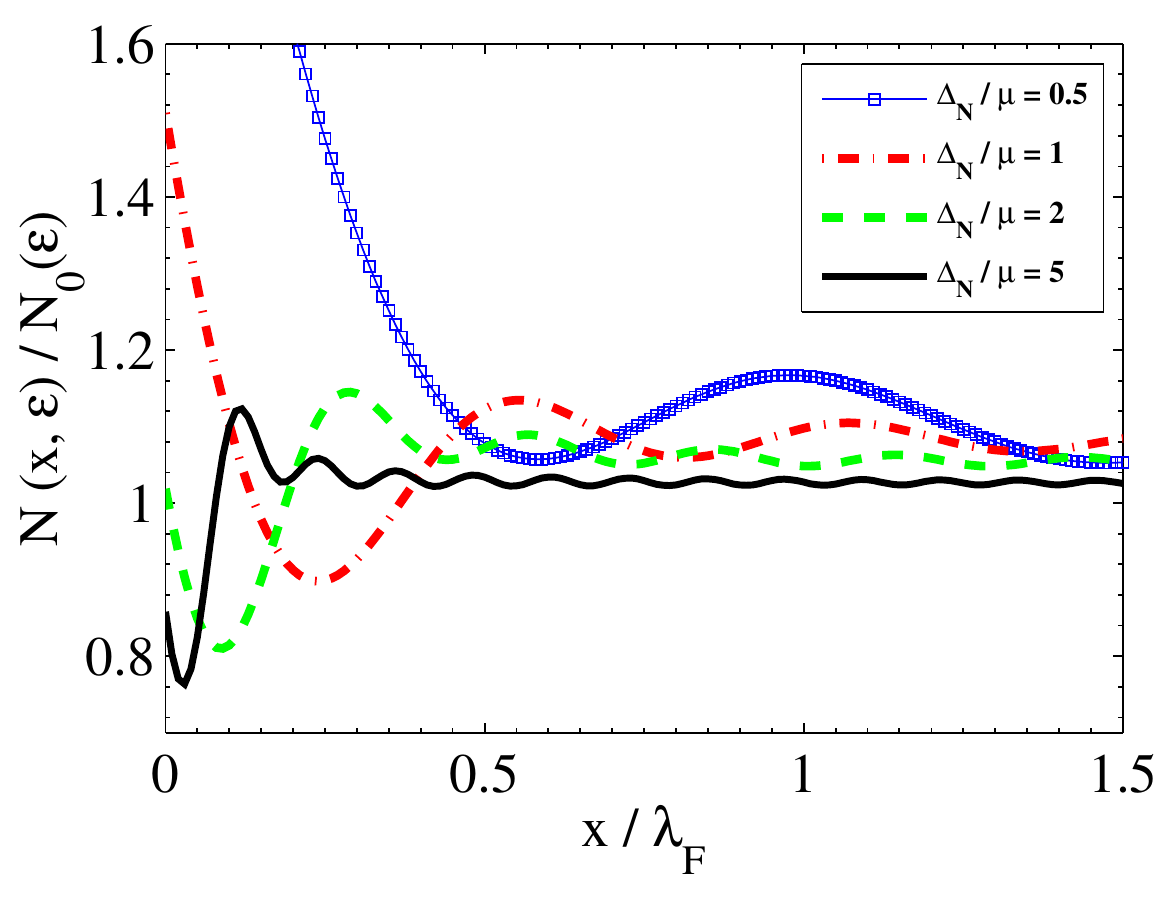}
\end{center}
\caption{\label{Fig:4}(Color online) The behavior of the proximity density of states (DOS) inside the PF region versus $x/\lambda_F$ for different values of $\Delta_N/\mu$, when $\varepsilon/\Delta_S = 0.5$ and $\Delta_N/\Delta_S = 0.1$. }
\end{figure}
Fig. \ref{Fig:4} shows the behavior of the proximity DOS inside the PF region in terms of the dimensionless distance $x/\lambda_F$ ($\lambda_F=\hbar v_F/\mu$) for different values of $\Delta_N/\mu$, when $\varepsilon/\Delta_S = 0.5$ and $\Delta_N/\Delta_S = 0.1 $. The results are for the case of AR with $\langle{\sigma}_z\rangle$ inversion at S/PF interface.
It is seen that $N(x)$ decays rather quickly close to the interface with a slope which increases by increasing $\Delta_N/\mu$. We can see that the length scale characterizing this fast decay decreases by increasing $\Delta_N/\mu$. Slightly away from the interface, a much slower oscillating behavior can be seen with the wavelength which clearly decreases with $\Delta_N/\mu$. Also the length scale which characterizes these oscillations decreases with $\Delta_N/\mu$.
\par
So we conclude that there are two phenomena to consider in describing the spatial variations of $N(x)$,
when an energy gap is present in the band structure of normal graphene. The first is the short distance decay at the interface and the other important   phenomenon is the damped oscillation of $N(x)$, caused by the momentum shift between Andreev correlated electron-hole with opposite $\langle{\sigma}_z\rangle$ directions. As can be seen from Fig. \ref{Fig:4}, the period of oscillations is determined by $\hbar v_F/\Delta_N$, which is similar to a S/F structure where the period of DOS oscillations in the ballistic limit is given $\hbar v_F/h$ \cite{Zareyan01,Zareyan02,Buzdin05}. This shows the similarity of the effect of an spin-splitting exchange field $h$ with the energy gap $\Delta_N$, which behaves as a pseudospin-splitting field [see Eq. (\ref{H})]. The general method to detect these oscillations is tunneling spectroscopy, which is widely used to probe the DOS oscillations in S/F structures with different thicknesses of the F layer [see e.g. Ref. \onlinecite{Kontos01}]. In our case, the local scanning of the surface of the PF region by scanning tunneling microscopy (STM) is the appropriate method to probe the DOS oscillations. We note that the imperfectness of the surface at S/PF junction cannot change the qualitative behavior of the results. It may lead to weakening the proximity effect and changing the results quantitatively\cite{Zareyan01,Moghaddam10}.
So the spatially-damped oscillatory behavior of the DOS inside the PF region confirms that the energy gap $\Delta_N$ in the band structure of normal graphene
 produces an effect similar to the exchange field in F graphene.
\section{\label{sec:level4}Crossed Andreev reflection in PF/S/PF structure}
We further study the non-local quantum transport and CAR in PF/S/PF junction which constitutes a superconducting pseudospin valve structure. In CAR process an electron excitation and a hole excitation at two separate PF leads are coupled by means of Andreev scattering processes at two spatially distinct interfaces. The superconducting pseudospin valve consists of two PF regions with a tunable direction of PM, which are connected through a S region of length $L$. The configuration of PMs in the pseudospin valve can be changed from parallel (P) to antiparallel (AP) by fixing the type of doping of one region and changing the type of the doping in the other region. Using the solutions of DBdG equation for the quasiparticles of PF and S regions, we write the wave functions inside the two PF and S regions of P and AP configurations, within the scattering formalism. Matching the wave functions at the two interfaces, we calculate the normal and Andreev reflection amplitudes in the left PF region and the transmission amplitudes of the electron and the hole into the right PF region of both P and AP configurations. Replacing the reflection and transmission amplitudes in BTK  formula, we obtain the conductance of AR, CT, and CAR processes for P and AP alignments of PMs. We find that for all incoming waves with two bias voltages $eV = \pm (\mu-\Delta_N)$, AR process is suppressed and the cross-conductance in the right PF region depends crucially on the configuration of PMs in the two PF regions. We find that the transport is mediated purely by CT in P configuration and changes to the pure CAR in the low energy regime, by reversing the direction of PM in the right PF region. This suggests a pseudospin switching effect between the pure CT and pure CAR in PF/S/PF structure, which can be seen from Eq. (\ref{pseudospin_e}) ( Eq.  (\ref{pseudospin_h})) for the right going conduction- (valance-)band electron (hole) of $n$- ($p$-)doped PF region by replacing $-\cos{\alpha_e}$ with $\cos{\alpha_e}$ ($\mu$ with $-\mu$).
\begin{figure}
\begin{center}
\includegraphics[width=3.4in]{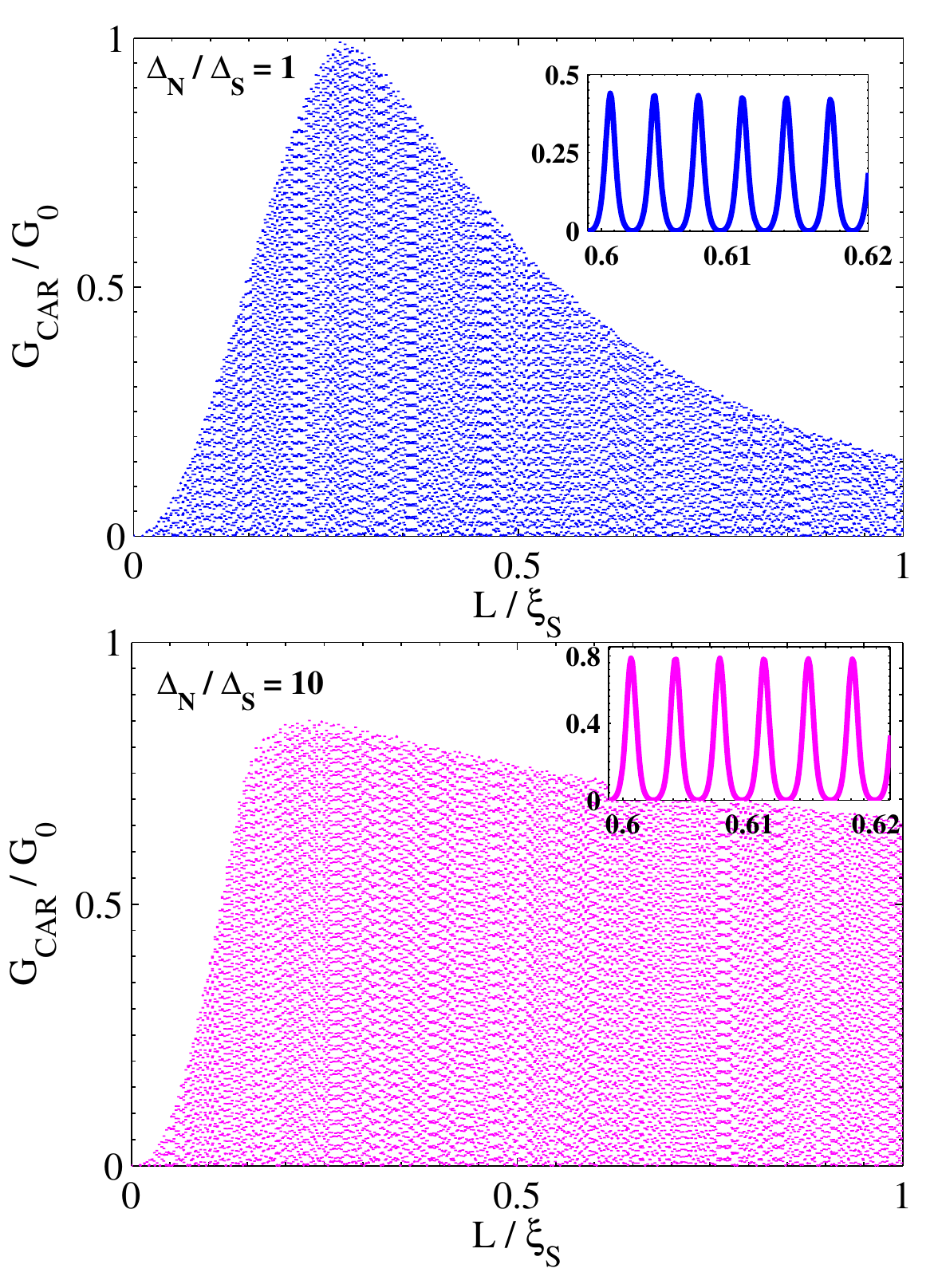}
\end{center}
\caption{\label{Fig:5}(Color online) Plots of the conductance of CAR process in antiparallel configuration of pseudomagnetizations versus the length of the S region for different two values of $\Delta_N/\Delta_S$, when $\mu/\Delta_N= 1.1$ and $eV = \mu - \Delta_N$.}
\end{figure}
\par
Fig. \ref{Fig:5} shows the behavior of the conductance of CAR process in AP configuration versus the length of the S region for different values of $\Delta_N/\Delta_S$, when $\mu / \Delta_N = 1.1$ and $eV = \mu - \Delta_N$. It is seen that the CAR conductance has an oscillatory behavior with $L / \xi$ and increases by increasing $\Delta_N/\Delta_S$ from its value for the corresponding graphene N/S/N structure. Also we can see that in contrast to the graphene N/S/N structure, CAR process is present for long lengths of the S region.
This effect is similar to graphene F/S/F structure \cite{linder09} and approves that the gapped normal graphene behaves like a F graphene.
\section{\label{sec:level5}Conclusion}
In conclusion, we have investigated proximity effect in graphene-based hybrid structures of superconductors and gapped regions as pseudoferromagnets. A gapped graphene is in a sublattice pseudospin symmetry-broken state with a net pseudomagnetization oriented perpendicularly to the plane of graphene. We have found that upon a certain condition, Andreev reflection of an electron from a S/PF interface is associated with an inversion of the perpendicular component of its pseudospin, and that this has important consequences for the proximity effect. For a S/PF junction, we have found that, the Andreev-Klein reflection can enhance the pseudospin inverted Andreev conductance by the energy gap $\Delta_N$ to reach a limiting maximum value for $\Delta_N\gg \mu$, which depends on the bias voltage and can be larger than the value for the corresponding junction with no energy gap ($\Delta_N\ll \mu$). This is similar to the behavior of Andreev conductance with the exchange energy $h$ in a graphene ferromagnet-superconductor junction. We have further studied the proximity density of states (DOS) in PF side of S/PF contact, which exhibit a damped-oscillatory behavior as a function of the distance from the interface. The period of DOS oscillations is found to be inversely proportional to the energy gap $\Delta_N$. The proximity DOS in ferromagnetic graphene shows similar spatial oscillations with a period determined by $1/h$. For a PF/S/PF structure, we have found a pseudospin switching effect in which the subgap transport of electrons can be switched from a purely elastic electron co-tunneling process to the pure crossed Andreev reflection by changing the alignment of the pseudomagnetizations of PF regions from parallel to antiparallel configurations. This is again similar to the behavior of the corresponding superconducting structure with ferromagnetic graphene. This confirms that, in this respect, the effect of the sublattice pseudospin degree of freedom in gapped graphene is as important as the spin in a ferromagnetic graphene.

\begin{acknowledgments}
The authors gratefully acknowledge support by the Institute for Advanced Studies
in Basic Sciences (IASBS) Research Council under grant No. G2010IASBS110.
\end{acknowledgments}

\end{document}